# High-rate continuous-variable quantum key distribution coexisting with Tb/s coherent classical transmission in hollow-core fiber

Xitao Ji[1,†], Siyu Chen[1,†], Peng Li[2], Mingming Zhang[1], Yilun Chen[1], Jun Gao[1,5,6], Rui Lin[3], Davide Bacco[4], Siqi Yan[1,5,6*] and Ming Tang[1,5,6*]

[1]School of Optical and Electronic Information, Huazhong University of Science and Technology, Wuhan 430074, China

[2]State Key Laboratory of Optical Fiber and Cable Manufacture Technology, Yangtze Optica Fibre and Cable Joint Stock Limited Company (YOFC), Wuhan 430074, China

[3]Department of Electrical Engineering, Chalmers University of Technology, Gothenburg 41296, Sweden

[4]Department of Physics and Astronomy, University of Florence, Sesto Fiorentino 50019, Italy

[5]Hubei Optical Fundamental Research Centre, Wuhan 430074, China

[6]Research Center for Cross-Scale Global-Scale Multimodal Optical Communication Discipline and Technology, Wuhan 430074, China

*Email: siqya@hust.edu.cn and tangming@hust.edu.cn

†These authors contributed equally to this work.

**Abstract**

Quantum key distribution (QKD) can provide secret keys with security rooted in quantum mechanics, but operation alongside high-capacity classical traffic remains limited by the excess-noise budget of weak quantum states in conventional solid-core fiber. Here, we combine ultralow-loss anti-resonant hollow-core fiber with residual-carrier-assisted discrete-modulation continuous-variable QKD (DM-CV-QKD) to address both propagation-induced coexistence noise and low-SNR phase recovery. Over a 24.3-km hollow-core link with 3.3-dB end-to-end loss, a dual-polarization 15-Gbaud DM-CV-QKD channel achieves an average asymptotic secret-key rate (SKR) of 153.22 Mb/s and a finite-size SKR of 149.99 Mb/s, while 37 coherent wavelength-division-multiplexed channels deliver 7.2 Tb/s net classical throughput. The system can even sustain a positive SKR under a high classical launch power of up to 15 dBm, without an optical bandpass filter (BPF). Finite-size analysis against collective attacks further yields a projected positive secret-key rate at a 100-km-equivalent condition. These results show that an anti-resonant hollow-core fiber, combined with carrier-assisted phase recovery, can greatly extend the operating regime of shared-fiber quantum-secured coherent links, pointing to a promising approach for integrating high-rate CV-QKD with high-capacity optical networks.

## Introduction

Quantum key distribution (QKD) can establish secret keys whose security is grounded in the laws of quantum mechanics rather than in computational hardness assumptions.[1] For QKD to move from dedicated experimental links to scalable quantum-secured networks, a central requirement is compatibility with the optical-fiber infrastructure already used for high-capacity data transmission.[2–4] In such a setting, the generated keys can be continuously supplied to high-throughput data-encryption systems, providing a practical route towards secure communication without relying on public-key cryptography for long-term confidentiality.[5–7] Among the main QKD approaches, continuous-variable QKD (CV-QKD) is particularly attractive because it encodes information in optical-field quadratures and can exploit coherent detection, high-speed modulators, and digital signal processing developed for classical telecommunications. These features make CV-QKD a natural candidate for metropolitan and access-scale links where high SKR, compact hardware, and network compatibility are simultaneously required.[8–10]

Shared-fiber CV-QKD, however, remains constrained by the coexistence of weak quantum states and much stronger classical optical channels. In conventional solid-core single-mode fiber (SMF), classical channels generate spontaneous Raman-scattering photons through light–glass interaction. Broadband Raman noise can overlap the quantum band and directly increase the excess noise, while fiber attenuation reduces the received quantum-state amplitude and therefore the noise margin available for secure key extraction. Recent demonstrations have pushed CV-QKD towards high-rate operation over 100-km-class channels and have shown coexistence with classical data in solid-core fiber links.[10,11] Dense-WDM studies have also clarified how wavelength allocation, channel spacing, and launch power affect the coexistence window.[12] These advances show the practical relevance of shared-fiber CV-QKD, but the underlying propagation medium remains silica core, where Raman scattering, nonlinear interaction, and attenuation jointly restrict the launch-power, distance, and SKR operating range.

Hollow-core fiber offers a direct way to relax this physical-layer constraint. Because the optical field propagates predominantly in air rather than silica, anti-resonant hollow-core fibers strongly reduce light-glass interaction and suppress Raman scattering and Kerr nonlinearity while maintaining low attenuation.[13–15] For CV-QKD, the value of hollow-core guidance becomes most pronounced in the coexistence regime. When weak quantum states propagate alongside much stronger coherent WDM channels, the secure operating margin is governed not only by link loss but also by how efficiently the classical fields generate excess noise in the fiber. By reducing the overlap between the optical field and silica glass, hollow-core fiber suppresses this noise-generation pathway at its origin, thereby easing the Raman- and nonlinearity-induced penalties that otherwise constrain the launch-power and key-rate window.

In this work, we demonstrate high-rate discrete-modulation CV-QKD co-propagating with high-capacity coherent classical transmission over a 24.3-km anti-resonant hollow-core fiber link with 3.3 dB end-to-end loss. The quantum subsystem uses probabilistically shaped, dual-polarization 16-QAM modulation at 15 Gbaud, together with residual-carrier-assisted frequency-offset and phase-noise compensation. In contrast to conventional pilot-assisted schemes that allocate separate temporal, spectral, or polarization resources, the residual carrier is derived from the same laser as the quantum signal and co-propagates through the same optical path, providing an intrinsic phase reference for low-SNR quantum-state demodulation. The classical subsystem consists of 39 wavelength-division-multiplexed, dual-polarization 16-QAM channels at 24.5 Gbaud, delivering 7.6 Tb/s gross and 7.2 Tb/s net throughput. Without narrowband optical filtering, the system achieves an asymptotic secret-key rate of 122.83 Mb/s at a total classical launch power of 0 dBm. With band-pass filtering, it sustains 77.28 Mb/s at 10 dBm. Rate-adaptive multi-edge-type low-density parity-check codes are used to support efficient reconciliation over the investigated loss range.[16–18] Beyond the coexistence demonstration, we quantify forward and backward spontaneous Raman scattering in the hollow-core link and examine how the secret-key rate depends on classical loading and quantum-channel wavelength. Using experimentally measured channel parameters obtained from the $10^7$-symbol experimental blocks, the finite-size secret-key rates against collective attacks are evaluated by extrapolating the security bounds to an expected practical block length of $10^{10}$ symbols, yielding a projected positive SKR up to a 100-km-equivalent transmission condition. These results show that hollow-core transmission, when combined with residual-carrier-assisted phase recovery, can enlarge the operating space in which high secret-key rates and high-capacity coherent transmission are simultaneously achievable.

## Results

## Transmission and Raman-noise characteristics of the AR-HCF

We first examined the structural and transmission characteristics of the anti-resonant hollow-core fiber (AR-HCF), which determine the propagation loss and fiber-generated noise relevant to quantum-classical coexistence. As shown in Fig. 1(a), the cladding comprises a single ring of non-touching silica capillaries surrounding an air-filled core. Anti-resonant reflection at the thin silica membranes confines the guided mode predominantly to the core, resulting in a small modal overlap with glass.[19] This reduced overlap directly weakens the light-glass interaction that gives rise to Raman scattering and Kerr nonlinearity in conventional solid-core fibers.

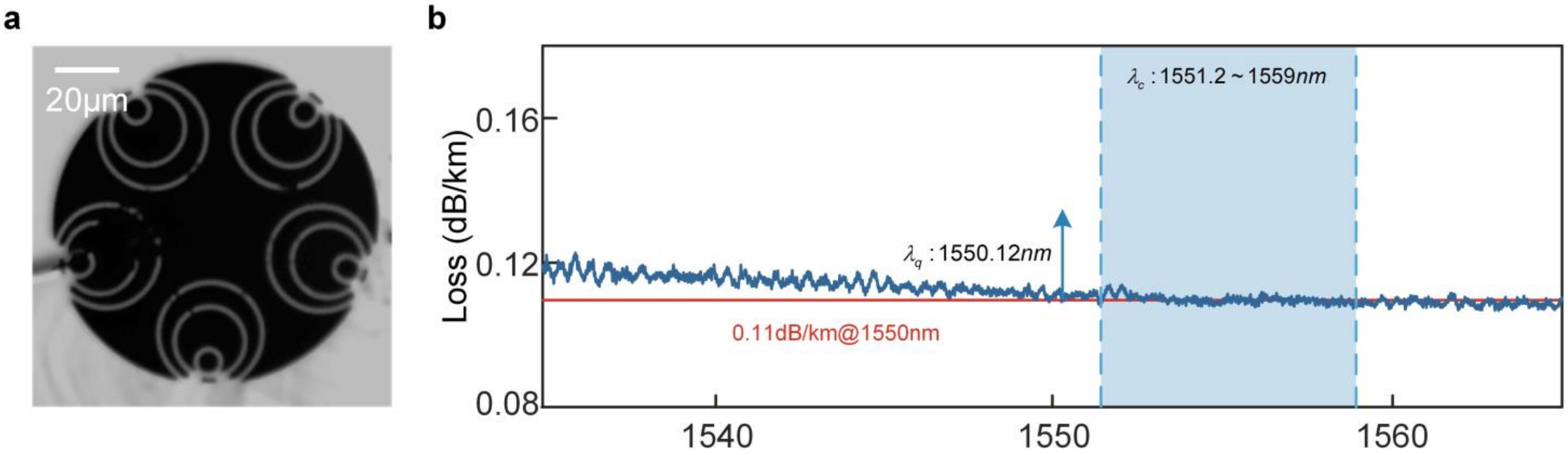

**Fig. 1 | Structural and transmission characteristics of the hollow-core link.** (**a**) Optical micrograph of the AR-HCF cross-section. (**b**) Measured attenuation spectrum of the 24.3-km HCF link.

The measured attenuation spectrum of the 24.3-km link is shown in Fig. 1(b). The C-band transmission window is free of pronounced absorption resonances and exhibits an attenuation of 0.11 dB km$^{-1}$ at the 1550 nm quantum wavelength. The minimum measured attenuation is approximately 0.10 dB km$^{-1}$ near 1565 nm. These measurement results demonstrate that the AR-HCF platform provides an ultra-low-loss and spectrally flat transmission window across the entire C-band, thereby validating the inherent suitability of this fiber infrastructure for high-capacity classical traffic and quantum signal coexistence.

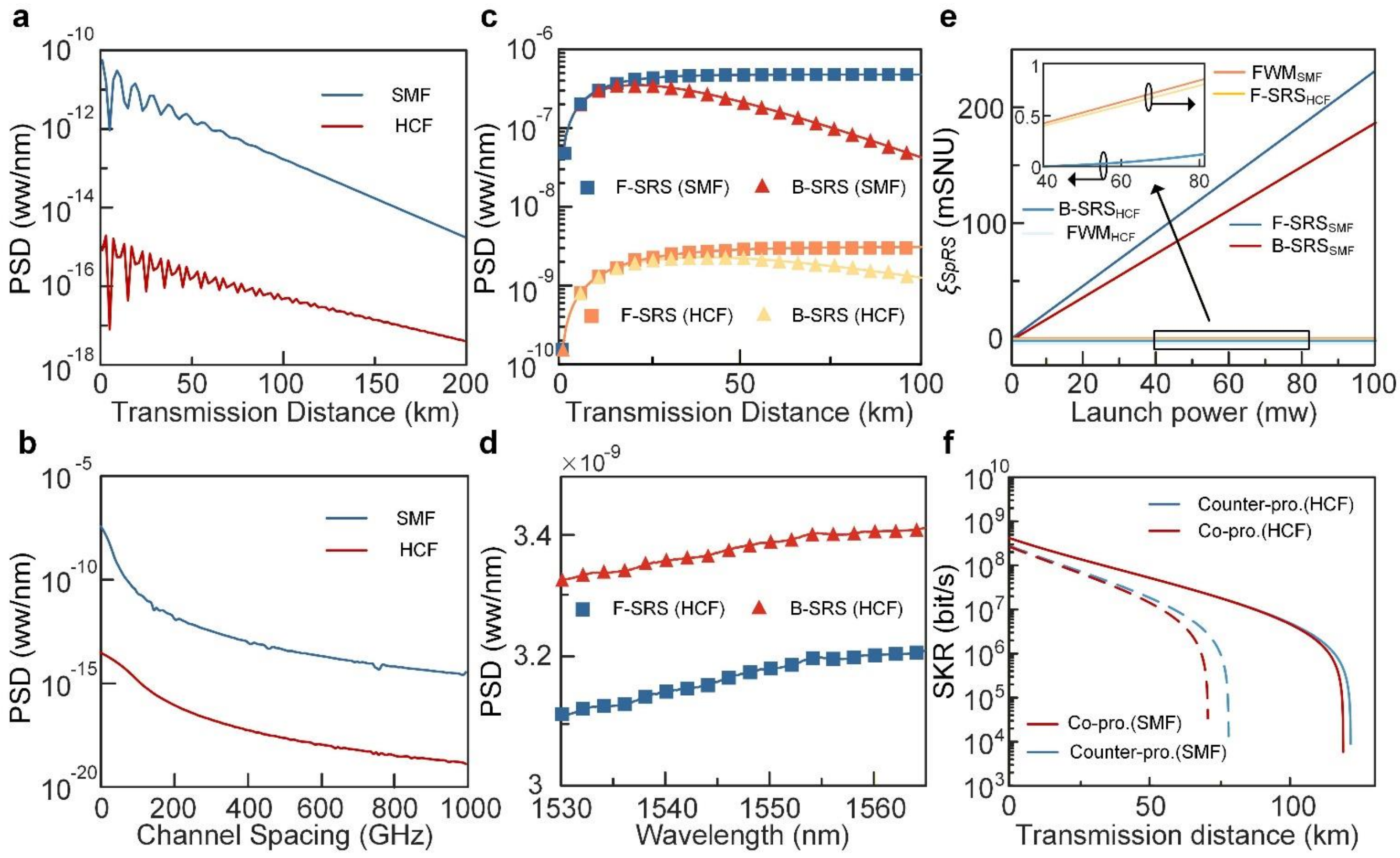

**Fig. 2 | Measurement and analysis of spontaneous Raman-scattering (SpRS) noise and SKR. (a)** FWM noise PSD versus transmission distance in SMF and HCF. **(b)** F-SRS and B-SRS noise PSD versus channel spacing in SMF and HCF. **(c)** F-SRS and B-SRS noise PSD versus transmission distance in SMF and HCF. **(d)** SpRS noise PSD in HCF versus quantum signal wavelength under a fixed classical wavelength condition. **(e)** Excess noise induced by FWM and SpRS noise versus total classical launch power. **(f)** SKR versus transmission distance for SMF and HCF platforms under co-propagating (Co-pro.) and counter-propagating (Counter-pro.) configurations.

Fig. 2 illustrates the nonlinear-noise characteristics relevant to QKD-classical coexistence in conventional SMFs and HCFs. Attention is initially directed toward four-wave mixing (FWM), which represents a coherent nonlinear process through which high-power wavelength-division-multiplexed classical channels can generate in-band noise near the quantum channel. The power spectral densities (PSDs) of this FWM noise as a function of transmission distance and channel spacing are displayed in Fig. 2(a) and (b), respectively. The physical evolution of the FWM noise power, denoted as $P_{FWM}$, is mathematically governed by[20]:

$$P_{FWM}(z) = \frac{\eta D^2 \gamma^2 P_i P_j P_k e^{-\alpha z}}{9\alpha^2}[1 - e^{-\alpha z}]^2 \quad (1)$$

where $P_i$, $P_j$, and $P_k$ denote the power levels of the three input channels, with $\alpha$ representing the fibre attenuation coefficient, $z$ representing the propagation distance, and $\gamma$ is the fibre nonlinearity parameter. $D$ is the FWM degeneracy factor. In conventional SMFs, the FWM noise power experiences spatial fluctuations and maintains a relatively high baseline over long transmission distances, whereas the HCF platform exhibits a significant reduction in FWM accumulation by several orders of magnitude due to its intrinsically small nonlinear coefficient $\gamma$. In addition to suppressing Kerr-induced nonlinear mixing, the reduced overlap between the optical field and silica glass in the AR-HCF also lowers the spontaneous Raman-scattering (SpRS) background. Fig. 2 (c) compares the measured forward and backward Raman-noise spectra of SMF and AR-HCF under identical launch conditions. The collected SpRS noise power $P_{SpRS}$ is defined as[3]:

$$P_{SpRS} = \begin{cases} \dfrac{P_0\beta}{\alpha_q - \alpha_c}[\exp(-\alpha_c L) - \exp(-\alpha_q L)], F-SRS \\ \dfrac{P_0\beta}{\alpha_q + \alpha_c}\{1 - \exp[(\alpha_q + \alpha_c)L]\}, B-SRS \end{cases} \quad (2)$$

Where $P_0$ is the classical pump power, $\beta$ is the Raman scattering cross-section, and $\alpha_q$, $\alpha_c$ represent the attenuation coefficient for the quantum and classical channels. Based on values reported in the literature, the $\beta$ of the SMF is $6 \times 10^{-9}$ $mW^{-1}nm^{-1}$, and $2.43 \times 10^{-9}$ $mW^{-1}nm^{-1}$ for the hollow-core architecture[21]. As shown in Fig. 2(c), both forward SpRS (F-SRS) and backward SpRS (B-SRS) noise in the SMF rapidly accumulate and saturate within the first 10 km of transmission. Conversely, the HCF platform achieves a suppression of SpRS noise by approximately three orders of magnitude, exhibiting a strictly linear growth for F-SRS and a gradual saturation profile for B-SRS at much lower absolute power levels.

The selection of the quantum wavelength in the coexistence framework involves a trade-off between Raman-noise suppression and fiber attenuation profile. For the WDM classical band utilized in this work, shifting the quantum channel toward shorter wavelengths reduces the spontaneous Raman scattering (SpRS) noise power spectral density (PSD) generated by the longer-wavelength classical channels, yet simultaneously shifts the quantum signal closer to the higher-loss edge of the anti-resonant hollow-core fiber (AR-HCF) transmission window. As illustrated in the experimental characterization across the C-band range over the 24.3-km link in Fig. 2(d), the SpRS noise PSD within the AR-HCF exhibits a distinct wavelength dependence governed by the joint profiles of fiber attenuation and Raman scattering. To optimize this trade-off while satisfying hardware implementation requirements, the quantum channel wavelength is strategically set at 1550.0 nm utilizing a high-coherence narrow-linewidth laser, while 39 coherent classical channels occupy the 1551.2 to 1559.0 nm spectral range. This grid allocation successfully places both the quantum and classical channels within the absolute minimum-loss window of the AR-HCF, maintaining sufficient spectral isolation between the quantum channel and the high-capacity coherent classical transmission band.

To assess how fiber-generated noise translates into the quantum-channel noise budget, Fig. 2(e) shows the calculated FWM and SpRS contributions as a function of the aggregate classical launch power. The excess noise is analytically quantified by:[12,20]

$$\xi_n = \frac{S_n(\lambda)\Delta\lambda_{eff}}{2R_s(\frac{hc}{\lambda_0})} \quad (1)$$

Here, $S_n(\lambda)$ represents the PSD, $\Delta\lambda_{eff}$ represents the effective optical bandwidth, and $R_s$ is the symbol rate. In the SMF, both F-SRS and B-SRS noise scale intensely with the launch power, creating a severe noise background that drives $\xi$ well beyond the secure operating range. By contrast, the HCF strongly suppresses both FWM and SpRS contributions over the same launch-power range. Even at a total classical launch power of 100 mW, the resulting excess noise remains below 1 mSNU, indicating that the hollow-core link provides a substantially larger noise margin for quantum-classical coexistence.

Finally, the ultimate system-level performance is evaluated through the simulated secure key rate (SKR) versus transmission distance under different classical-quantum co-propagation schemes, as illustrated in Fig. 2(f). For the SMF platform, the severe accumulation of non-linear noise restricts the maximum transmission distance to less than 80 km, where the counter-propagating configuration (Counter-pro.) slightly outperforms the co-propagating scheme (Co-pro.) due to the reduction in forward scattering noise. For the HCF platform, owing to the ultra-low FWM and SpRS noise profiles, the secure transmission distance is successfully extended to 120 km. Moreover, the HCF demonstrates high resilience to non-linear impairments, maintaining a multi-megabit-per-second SKR over practical distances and narrowing the performance gap between the co-propagating and counter-propagating architectures.

**Carrier-assisted phase recovery for low-SNR quantum-state demodulation**

For high-rate DM-CV-QKD, reliable phase recovery is challenging because the quantum symbols are transmitted at low optical power and provide only a limited signal-to-noise ratio for estimating laser-frequency offset and phase noise. Conventional pilot-assisted approaches can improve phase estimation, but they typically require additional temporal, spectral, or polarization resources and may increase the transmitter dynamic range or DSP burden. Conventional pilot-assisted approaches can improve phase estimation, but they typically consume additional temporal, spectral, or polarization resources, thereby increasing the transmitter dynamic range or consumption of the digital signal processing (DSP). Residual-carrier modulation (RCM) provides a more compact alternative by intentionally biasing the in-phase/quadrature (IQ) modulator away from the carrier-suppression point, allowing a controlled fraction of the optical carrier to remain. The quantum sideband is then digitally shifted away from this residual carrier. Because the residual carrier and the quantum sideband originate from the same master laser and propagate through the identical optical path, the carrier naturally captures the common-mode frequency offset and phase fluctuations experienced by the quantum signal. Consequently, this residual carrier serves as an intrinsic, co-propagating phase reference for receiver-side frequency-offset estimation and phase-noise compensation without allocating an independent, separately modulated pilot channel[22]. The optical spectrum is illustrated in Fig. 3(a). Due to the finite extinction ratio of the practical modulator, a residual peak remains at the baseband center frequency even under the carrier-suppression modulation condition. The quantum signal is digitally up-converted and frequency-shifted by 11.25 GHz away from the carrier zone to avoid classical interference.

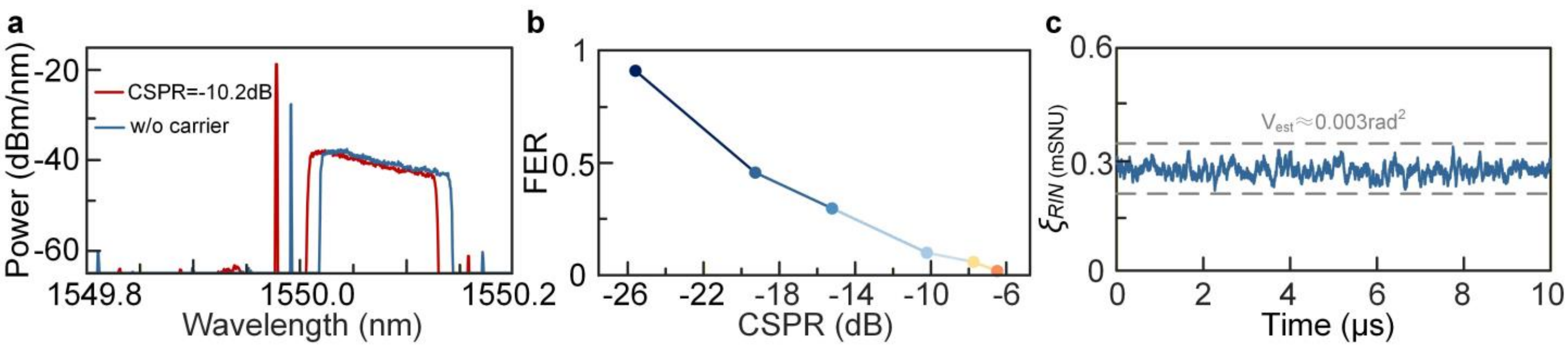

**Fig. 3 | Optimization of residual-carrier-assisted phase recovery in DM CV-QKD.** (**a**) Transmitted optical spectra with and without the residual carrier. (**b**) Measured FER versus CSPR. (**c**) Residual-phase-noise-induced excess noise

over a 10-μs acquisition interval at 24.3 km.

The quality of the residual carrier at the receiver directly determines the accuracy of frequency-offset and phase-noise estimation. A sufficiently strong carrier improves phase tracking, whereas excessive carrier power reduces the optical power allocated to the quantum sideband and degrades the effective quantum-signal SNR. The carrier-to-signal power ratio (CSPR) must therefore be optimized to balance phase-estimation accuracy against quantum-symbol quality.[23,24] As illustrated in Fig. 3(b), the measured frame-error rate (FER) is evaluated as a function of the CSPR under a fixed MET-LDPC code rate of R = 0.2. Due to the variations in the CSPR, the balance between carrier phase tracking precision and quantum-signal SNR is altered, leading to a dynamic fluctuation in the actual decoding FER. A CSPR of -10.2 dB provides an optimal operating point, yielding a baseline FER of 14.5%. At lower CSPR levels, insufficient carrier power limits phase estimation accuracy and introduces residual phase noise, whereas at higher CSPR levels, the diminished quantum-sideband power explicitly degrades the effective SNR, both cases resulting in an elevated reconciliation error rate.

To verify the suppression efficacy of the RCM technology against phase fluctuations, we investigate the excess noise specifically contributed by the residual phase noise $\xi_{RPN}$, which can be calculated as[25]:

$$\xi_{RPN} = TV_A(1 - e^{-V_{est}/2}) \tag{2}$$

where T is the channel transmittance, $V_A$ is the modulation variance of Alice's side, and $V_{est}$ is the residual phase-error variance. At the receiver, the residual carrier and quantum sideband are separated after coherent detection, and the carrier is used to estimate the frequency offset and phase noise experienced by the quantum signal. For the 24.3-km link, $V_{est}$ is approximately 0.003 rad$^2$, and $\xi_{RPN}$ remains near 0.3 mSNU over the measured interval, as shown in Fig. 3(c). As the equivalent distance increases, attenuation reduces the received carrier power and degrades phase estimation accuracy, leading to a gradual increase in residual phase noise.

**Experimental setup for quantum-classical coexistence over AR-HCF**

Fig. 4(a) shows the shared-fiber experimental architecture for co-propagating DM-CV-QKD and coherent classical transmission over the AR-HCF link. The quantum channel is located at 1550 nm, and 39 coherent classical channels occupy 1551.2–1559.0 nm on a 25-GHz grid. The multiplexed signals propagate through the same 24.3-km AR-HCF link and are separated at the receiver for independent coherent detection and offline DSP. Fig. 4(b) and (c) summarize the transmitter- and receiver-side processing chain used for the DM-CV-QKD channel, including residual-carrier generation, quantum-symbol modulation, carrier-assisted phase recovery, and reconciliation.

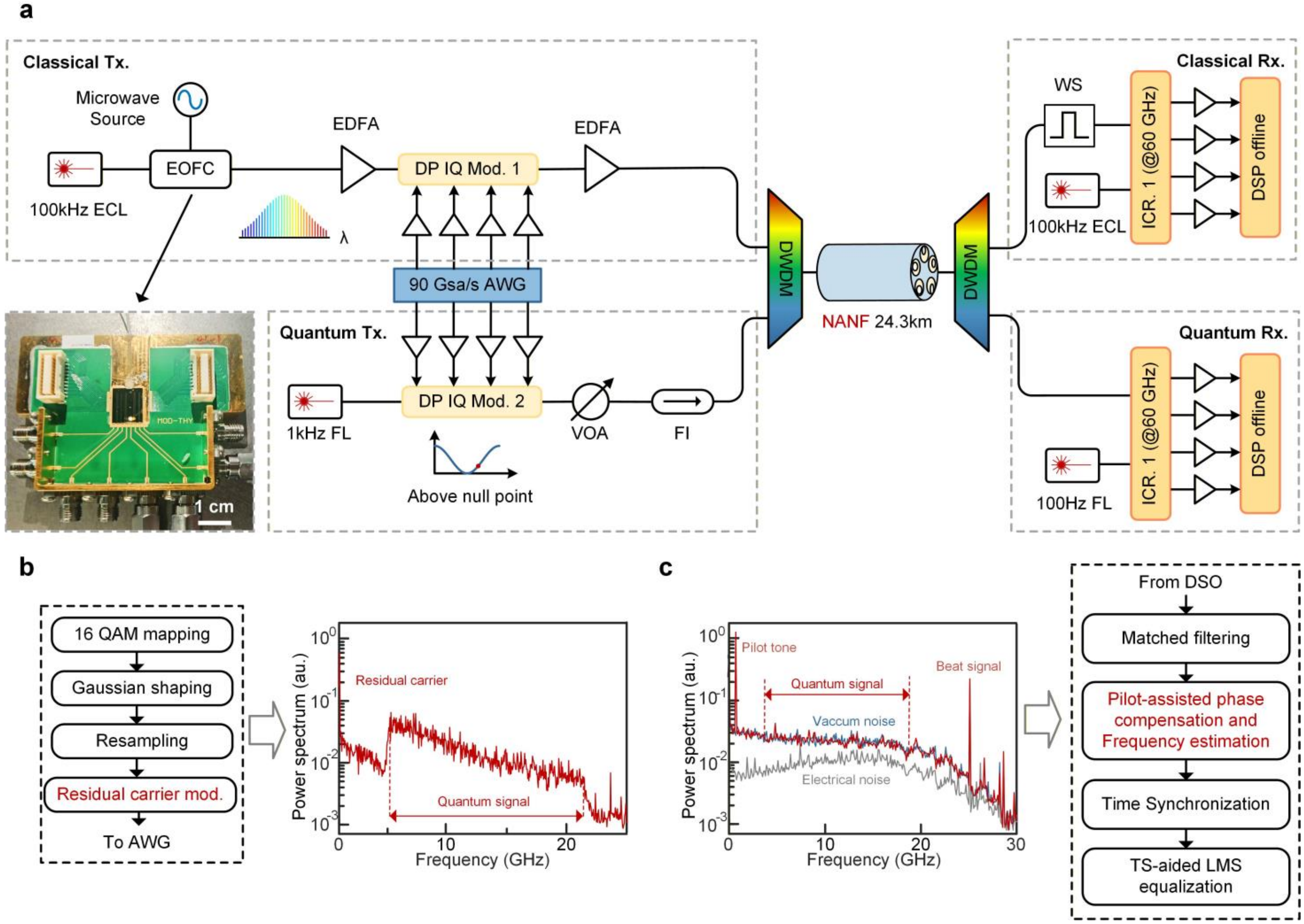

**Fig. 4 | Experimental setup for co-propagating coherent classical transmission and DM CV-QKD over AR-HCF.** **(a)** The classical channels are generated from an electro-optic frequency comb and modulated with dual-polarization probabilistically shaped 16-QAM. The quantum channel uses dual-polarization probabilistically shaped 16-QAM with a residual optical carrier. ECL, external-cavity laser; FL, fiber laser; EOFC, electro-optic frequency comb; DWDM, dense wavelength-division multiplexer; DP-IQ Mod, dual-polarization IQ modulator; EDFA, erbium-doped fiber amplifier; VOA, variable optical attenuator; FI, Faraday isolator; AR-HCF, anti-resonant hollow-core fiber; WS, wave shaper; ICR, integrated coherent receiver; DSO, digital storage oscilloscope; RTO, real-time oscilloscope. **b-c** DSP workflow for the high-rate DM CV-QKD system. **(b)** Transmitter DSP and transmitted spectrum. **(c)** Receiver DSP and detected spectrum.

For the classical subsystem, a 100-kHz-linewidth external-cavity laser (ECL; MTP-1000) with an output power of 13.9 dBm drives a thin-film lithium-niobate electro-optic comb generator comprising one intensity modulator and two cascaded phase modulators.[26] After polarization-maintaining erbium-doped fiber amplification, the comb has a total power of 9.8 dBm, 25-GHz line spacing, and a 10-dB spectrum width of 7.8 nm. A 90-GSa/s arbitrary waveform generator (Keysight M9502A) produces 24.5-Gbaud probabilistically shaped 16-QAM waveforms, which are applied to a dual-polarization IQ modulator to encode all comb lines simultaneously. The launch power is adjusted by an erbium-doped fiber amplifier. At the receiver, individual channels are selected by a programmable wave shaper with a 25-GHz passband, mixed with a local oscillator in an integrated coherent receiver (NeoPhotonics Class 60), sampled at 80 GSa/s, and processed offline.

For the QKD subsystem, a 1-kHz-linewidth fiber laser at Alice's side provides the optical carrier. In the implemented discrete-modulation protocol, probabilistically constellation-shaped 16-state quadrature amplitude modulation (PCS-16QAM) is selected as the primary modulation format for the system. To achieve frame synchronization, a joint cross-correlation is performed between the received complex fields and a local pulse-shaped template of the transmitted sequence. Before electro-optic conversion, the aggregate baseband quantum signal is shifted digitally to create sufficient spectral separation from the unshifted residual carrier component. This strategic arrangement allows the combined quantum sideband and the co-propagating carrier to be modulated simultaneously by a dedicated dual-polarization IQ modulator. A variable optical attenuator (VOA) is subsequently adjusted to set the modulation variance $V_A$ of the transmitted quantum states, followed by a Faraday isolator to suppress back-reflections and eliminate potential Trojan-horse security vulnerabilities.[27] quantum and classical signals are then wavelength-multiplexed and launched into the shared AR-HCF. At the receiver side, an independent laser provides a 13 dBm local oscillator (LO, NKT Koheras BASIK X15). To evade low-frequency noise impairments, the local oscillator is intentionally offset in the optical domain to beat with the unshifted residual carrier, generating an optical pilot tone at a 1-GHz intermediate frequency (IF). The mixed optical fields are detected by an integrated coherent receiver (Fujitsu FIM24725), and the four resulting electrical quadrature outputs are captured by a synchronized real-time oscilloscope running at a sampling rate of 50 GSa/s. To achieve high-precision phase recovery under intensive classical traffic, the digitized carrier is extracted via a pilot-matched digital filter; the quantum signal is then multiplied directly by the complex conjugate of the recovered carrier envelope, successfully compensating for both the residual frequency offset and high-frequency phase noise through a self-heterodyne-equivalent mechanism. Residual polarization mixing and dynamic channel fluctuations are subsequently compensated through a superimposed training-sequence-aided least mean square equalization (TS-aided LMS equalization) framework. In this architecture, identical training sequences are repeatedly superimposed in the time domain to function effectively as an additive white Gaussian noise (AWGN) filter, thereby enabling the LMS algorithm to robustly update its tap coefficients and track channel states under low-SNR conditions.[10] The comprehensive details of the transmitter and receiver DSP architectures are provided in Methods. Signal, shot-noise, and electronic-noise measurements are interleaved at 5-min intervals and processed using the same frozen DSP parameters. Due to the utilization of a power-stabilized local oscillator laser, the shot noise remains invariant within the calibration intervals, allowing the input-referred excess noise to be calculated as:[28]:

$$\xi = \frac{2(V_B - 1 - v_{ele})}{\eta T_{ch} T_{IL}} - V_A \tag{3}$$

where $T_{ch}$ is the channel transmittance, $T_{IL}$ is the insertion loss of the DEMUX and MUX, which is 1.2dB, and $\eta = 0.56$ is the measured receiver quantum efficiency.

Post-processing comprises reverse reconciliation with MET-LDPC codes followed by privacy amplification.[19–21,34] To accommodate the low and distance-dependent SNR, the reconciliation rate is adapted by puncturing and shortening Raptor-like parity-check matrices. The asymptotic SKR under collective attacks is evaluated using the Devetak–Winter expression:[29]

$$R_\infty = 2R_s R_c (1 - v_{TS})(1 - FER)(\beta I_{AB} - \chi_E) \tag{4}$$

where $R_s$ denotes the symbol rate, $R_c$ denotes the LDPC core rate, FER is the frame-error rate, and the factor of 2 accounts for the two polarization channels. For each transmission condition, $V_A$, code rate, reconciliation efficiency $\beta$, and FER are jointly optimized to maximize the SKR.

At a total classical launch power of 0 dBm and without an additional narrowband optical filter, the optimum operating point uses $V_A = 3.4$ SNU, channel SNR = −4.55 dB, and code rate R = 0.2, $\beta = 92\%$. Under these conditions, the 24.3-km link yields an SKR of 121.28 Mb/s and an average excess noise of 0.0202 SNU.

## Classical and quantum transmission performance

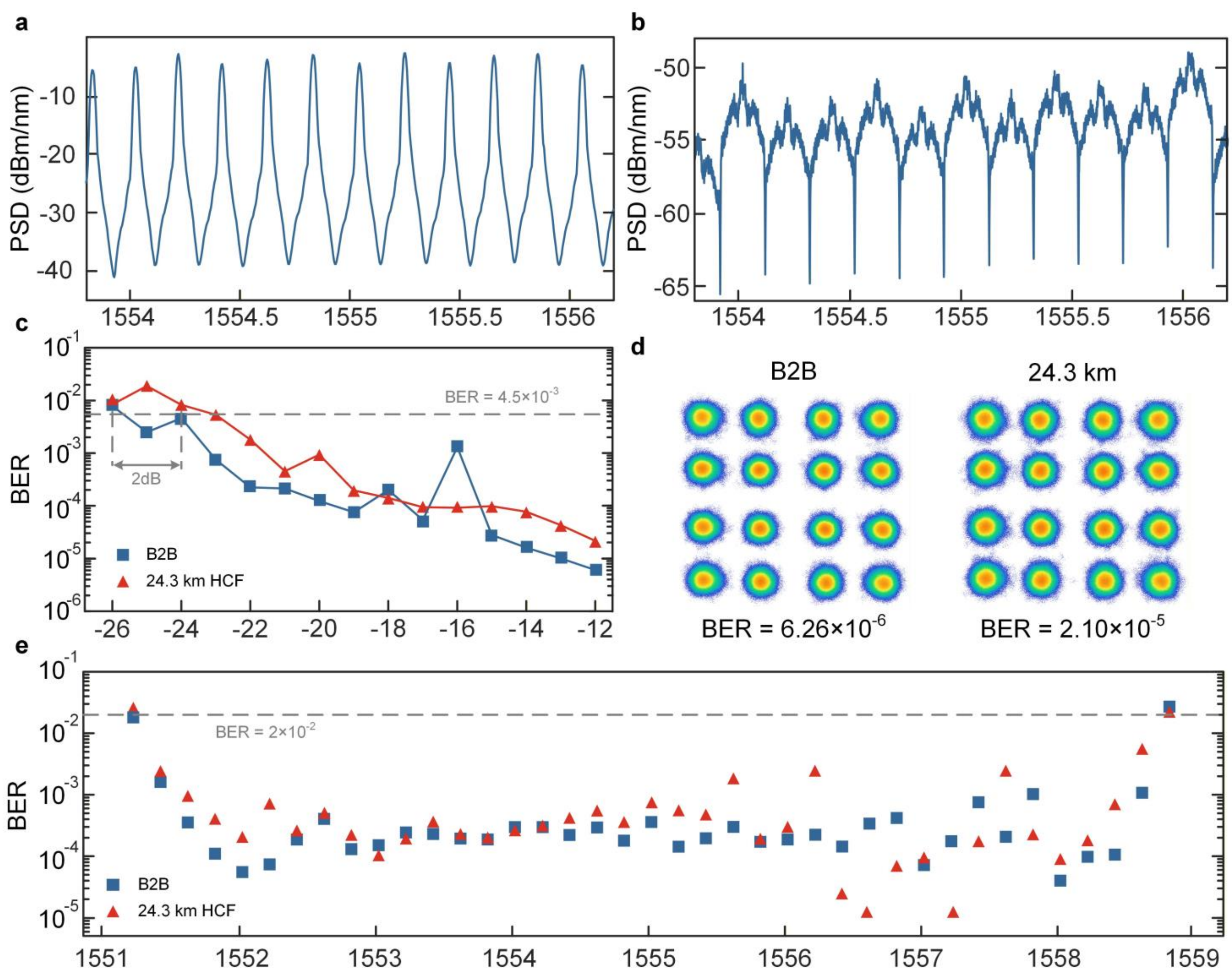

**Fig. 5 | WDM classical transmission using a broadband electro-optic comb.** Optical spectra of the unmodulated (**a**) and modulated (**b**) comb. (**c**) BER versus received optical power for back-to-back and 24.3-km HCF transmission. (**d**) Measured 16-QAM constellations at -12 dBm. (**e**) BERs of all WDM channels after transmission; 37 channels are below the soft-decision FEC threshold.

We demonstrate the performance of both quantum and classical signals within the AR-HCF to validate its capability as a single-channel medium for high-rate quantum-secured communications. Fig. 5(a) and (b) show a section of the optical spectra of the original and modulated electro-optic (EO) combs. As characterized in Fig. 5(a), the unmodulated optical frequency comb (OFC) exhibits sharp, discrete spectral peaks with high spectral purity, providing stable multi-wavelength carriers for high-capacity DWDM operations. Following data modulation, Fig. 5(b) depicts a distinct spectral evolution where the carriers transform into high-density, continuous modulation sidebands. For the classical channels operating under standard coherent formats, the optical power is distributed uniformly across the signal bandwidth, manifesting as flat spectrum envelopes without any residual carrier accumulation while maintaining sufficient guard bands. Fig. 5(c) illustrates the Bit Error Rate (BER) as a function of the received optical power (ROP) for the central 1555 nm channel under both back-to-back (B2B) and 24.3 km HCF transmission scenarios. While the B2B configuration consistently exhibits the lowest BER across the entire power range, indicating near-ideal system performance, the BER degradation after 24.3 km of HCF transmission remains minimal owing to the fiber's low-impairment characteristics. To achieve a BER below the forward error correction (FEC) threshold of $4.5\times10^{-3}$, the fiber link introduces a power penalty of 2 dB. Fig. 5(d) shows the constellation diagrams at an ROP of -12 dBm, with BERs of $6.26\times10^{-6}$ (B2B) and $2.10\times10^{-5}$ (24.3 km HCF), respectively.

The BERs of all 39 comb lines are summarized in Fig. 5(e). After transmission, 37 channels remain below the soft-decision FEC threshold of $2\times10^{-2}$. Each 24.5-Gbaud dual-polarization 16-QAM channel carries a gross rate of 196 Gb/s, corresponding to 7.6 Tb/s across all 39 channels. Counting the 37 channels below the FEC threshold gives a net supported throughput of 7.2 Tb/s. The small wavelength-dependent penalty is consistent with the measured link loss and transceiver response; no pronounced nonlinear degradation is observed over the investigated launch-power range.

Fig. 6(a) and (b) display the measured excess noise and SKR versus total classical launch power for coexistence operation with and without an optical band-pass filter (BPF), using $1 \times 10^7$ quantum symbols per measurement block. As illustrated in Fig. 6(a), as the total launch power of the classical channels increases from 0 to 15 dBm, it monotonically suppresses the SKR from 162.02 Mb/s to 36.27 Mb/s, and the excess noise increases from 0.0203 to 0.0526 SNU. When the launch power exceeds 15 dBm, the SKR drops to 0 because the excess noise exceeds the security threshold in the numerical simulation, preventing a positive secure key. These results confirm the system operation under a classical background power of up to 15 dBm due to the low SpRS coefficient of the HCF platform.

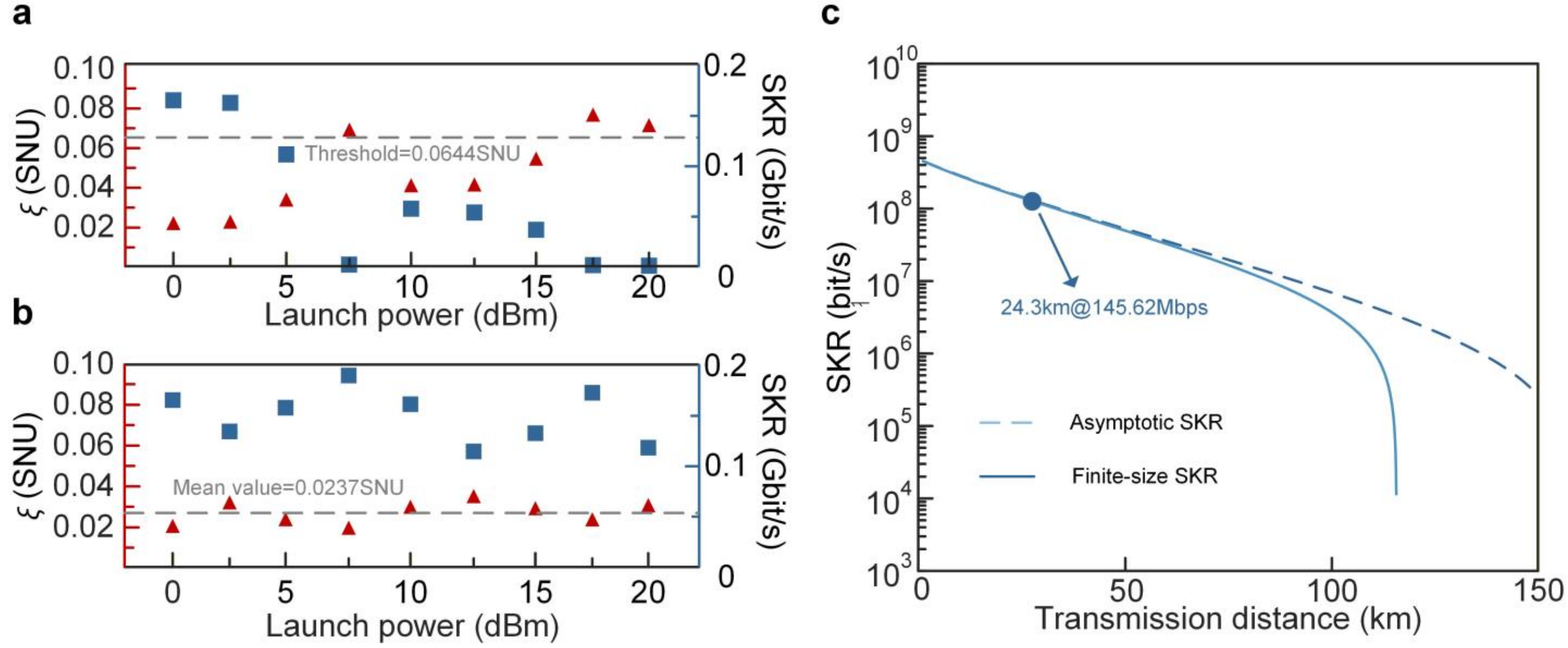

**Fig. 6 | Quantum-channel performance under classical coexistence and increasing loss.** Measured excess noise (triangles) and SKR (squares) versus total classical launch power with (**a**) and without (**b**) optical band-pass filtering. (**c**) Asymptotic and finite-size SKRs versus equivalent transmission distance; markers denote results evaluated from measured channel parameters at 24.3, 50, 75, and 100 km.

In Fig. 6(b), with the introduction of a BPF, the out-of-band SpRS noise and classical crosstalk are effectively suppressed. As the total classical launch power increases from 0 to approximately 20 dBm, the average channel excess noise is 0.0237 ± 0.004 SNU and the average asymptotic SKR of 145.62 Mb/s. The peak SKR of 186.41 Mb/s is achieved at a specific launch power of 7.5 dBm. The integration of the filter successfully extends the classical power margin while maintaining the SKR at the hundred-megabit-per-second level.

Fig. 6(c) plots the asymptotic and finite-size SKRs versus equivalent transmission distance. The 24.3-km point is measured on the physical HCF link; the 50-, 75-, and 100-km conditions are emulated by adding calibrated attenuation with a VOA. The measured channel parameters are then used to evaluate the corresponding SKRs. The system maintains positive asymptotic and finite-size SKRs throughout the investigated loss range, with the finite-size curve decreasing more rapidly as statistical uncertainty becomes dominant. Furthermore, owing to the low-loss characteristic of the hollow-core fiber, the transmission distance can theoretically be extended beyond 150 km in the asymptotic regime through the optimization of the modulation variance and the suppression of the excess noise.

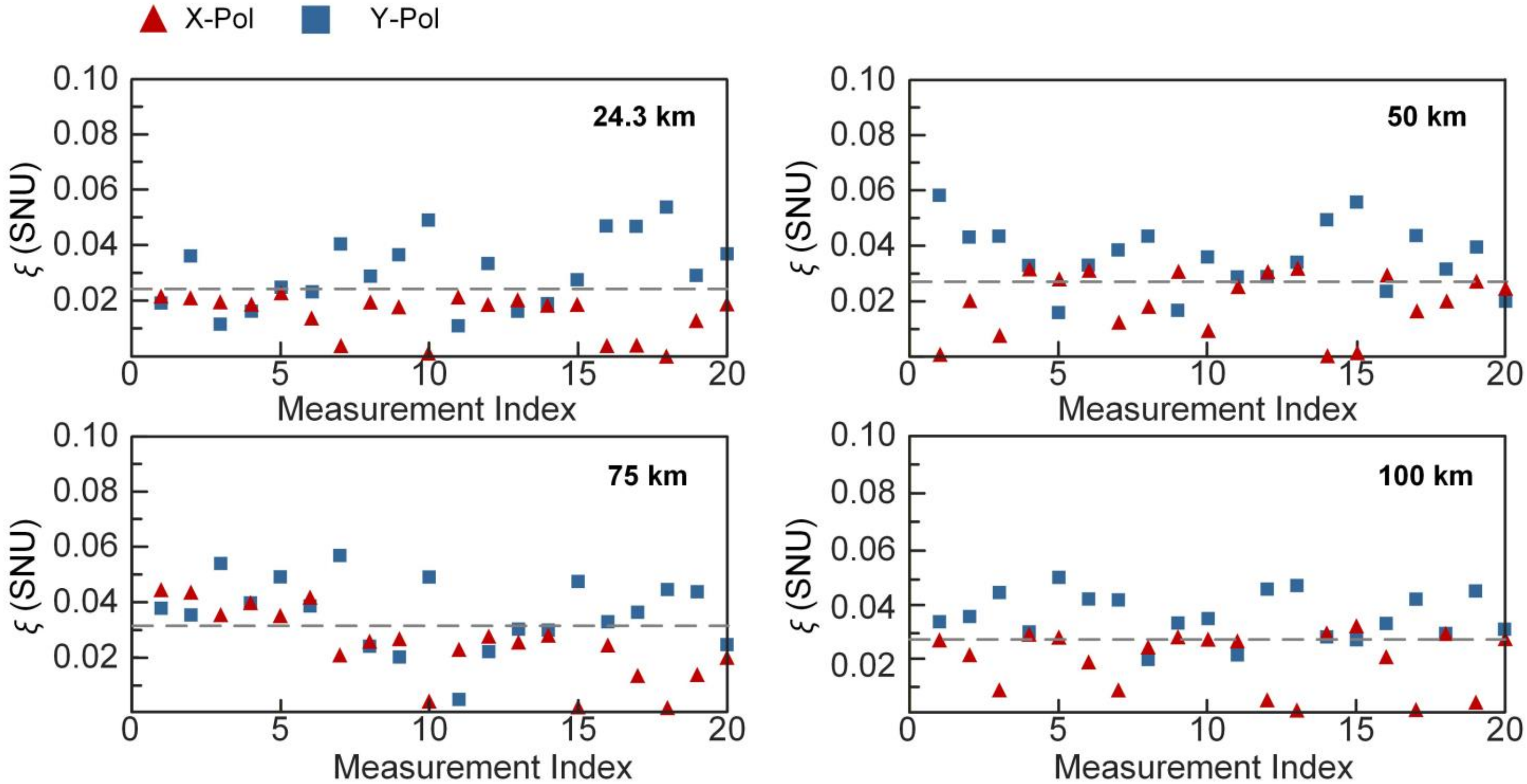

**Fig. 7 | Long-term excess-noise stability.** Excess noise measured over 20 measurements at the (a) 24.3 km, (b) 50 km, (c) 75 km, and (d) 100 km-equivalent transmission conditions. The gray dash lines denote the mean value of the excess noise, which is 0.0226 SNU@24.3-km, 0.0278 SNU@50-km, 0.0309 SNU@75-km, and 0.0279 SNU@100-km.

Finite-size parameter estimation is performed against collective attacks. For a parameter-estimation failure probability $\varepsilon_{PE} = 1 \times 10^{-10}$, the lower confidence bound on channel transmittance, $T^{low}$, and the upper confidence bound on excess noise, and the upper confidence bound on excess noise, $\xi_{up}$, are used to upper-bound Eve's Holevo information.[30,31] Each experimental excess-noise estimate uses $N_{sub} = 1 \times 10^{7}$ symbols, whereas the finite-size SKR projections assume a total block size $N = 1 \times 10^{10}$. The reported finite-size rates should therefore be interpreted as projections based on measured channel parameters rather than keys distilled from a complete $10^{10}$-symbol experimental block. Table 1 summarizes the corresponding operating parameters and the achievable asymptotic and finite-size SKRs, indicating that a projected positive finite-size SKR remains sustainable even at the 100-km equivalent transmission condition.

Beyond these discrete frame evaluations, the long-term operational robustness of the co-propagation architecture is systematically verified. The experimentally estimated excess noise ξ evaluated during a continuous 20-measurement operational window across different transmission distances is illustrated in Fig. 7. The temporal profiles confirm that the excess noise fluctuations remain well within the secure operational threshold throughout the entire tracking period, demonstrating excellent environmental immunity and confirming the system stability required for continuous-variable quantum key distribution infrastructures under high-power classical traffic.

| Symbol rate (Gbaud) | Distance (km) | Channel loss (dB) | $V_A$ (SNU) | $\xi$ (SNU) | Asymptotic SKR (Mb/s) | Finite-size SKR (Mb/s) |
|---|---|---|---|---|---|---|
| 15 | 24.3 | 3.3 | 3.0 | 0.0226 | 153.22 | 149.99 |
| 15 | 50 | 6.1 | 2.3 | 0.0278 | 21.36 | 19.98 |
| 15 | 75 | 8.9 | 1.9 | 0.0309 | 2.64 | 1.94 |
| 15 | 100 | 11.6 | 2.0 | 0.0279 | 0.37 | 0.06 |

**Table 1 | Measured channel parameters and calculated asymptotic and finite-size SKRs.**

**Discussion**

The results demonstrate that ultralow-loss AR-HCF can substantially enlarge the coexistence window of high-rate DM-CV-QKD and coherent classical transmission. The 24.3-km link combines 3.3 dB end-to-end loss with a Raman-scattering coefficient approximately two orders of magnitude below that used for SMF in the model. This combination directly improves the quantum-channel noise budget: low attenuation preserves the received quantum-state amplitude, while suppressed light–glass interaction reduces the Raman-noise floor generated by the classical channels. Consequently, 39 C-band coherent WDM channels can co-propagate with a 15-Gbaud dual-polarization quantum channel while the input-referred excess noise remains within the secure operating range. The simultaneous achievement of 7.2-Tb/s net classical throughput and 145.62 Mb/s SKR shows that, in a hollow-core link, classical capacity and quantum key rate are not forced into the same direct trade-off that typically constrains solid-core coexistence systems.

Residual-carrier-assisted phase recovery provides the complementary receiver-side mechanism required to use this enlarged noise margin. The co-propagating residual carrier tracks the common laser-frequency offset and phase fluctuations without assigning an additional time slot, polarization, or independently modulated digital pilot. Its cost is captured by the carrier-to-signal power ratio: insufficient carrier power degrades phase estimation, whereas excessive carrier power reduces the effective quantum-sideband SNR. At the selected CSPR of (-10.2) dB, the system supports 15-Gbaud dual-polarization operation together with rate-adaptive reconciliation. Optical band-pass filtering further increases the tolerable classical launch power by reducing out-of-band noise coupled into the quantum receiver, allowing positive high-rate key generation under stronger classical loading.

Table 2 compares the present work with representative quantum–classical coexistence demonstrations based on both discrete-variable and continuous-variable QKD protocols.[2,3,10–12,32] These prior studies have established important benchmarks in long-distance coexistence, high-rate CV-QKD, dense-WDM operation, and network-oriented resource allocation. The present architecture addresses a different physical-layer regime by combining hollow-core transmission with carrier-assisted DM-CV-QKD. Rather than relying only on wavelength planning or launch-power optimization within solid-core fiber, the AR-HCF link reduces the generation of coexistence noise at the propagation level, while the residual carrier stabilizes low-SNR quantum-state demodulation. This joint channel-and-receiver design enables high-capacity coherent classical transmission and high-rate secret-key generation over a single shared fiber, providing a scalable physical-layer approach for quantum-secured optical networks.

| | Classical rate | Secret-key rate | Distance | Loss | Protocol | Security analysis |
|---|---|---|---|---|---|---|
| This work | 7.2 Tb/s | 153.22 Mb/s (149.99 Mb/s finite-size) | 24.3 km | 3.3 dB | DM CV-QKD | Finite-size collective |
| Ref. 2 | 40 Tb/s | 9.56 kb/s | 101.6 km | 16.1 dB | DV-QKD | Finite-size collective |
| Ref. 3 | 110 Tb/s | 104 kb/s | 25.2 km | 17.9 dB | DV-QKD | Asymptotic collective |
| Ref. 10 | Not reported | 1,819.32 Mb/s | 5 km | 0.95 dB | GM CV-QKD | Finite-size collective |
| Ref. 12 | 1.6 Tb/s | 1.09 Mb/s | 10 km | 4.4 dB | GM CV-QKD | Asymptotic collective |
| Ref. 22 | 170 Gb/s | 43.2 kb/s | 120 km | 20 dB | GM CV-QKD | Finite-size collective |

**Table 2 | Comparison with representative quantum-classical coexistence demonstrations.**

Several points define the scope of the present demonstration and indicate directions for further development. First, the 50-, 75-, and 100-km-equivalent conditions are emulated using lumped attenuation rather than physically longer HCF spans. These measurements reproduce the loss budget, but do not capture distributed scattering, longitudinal modal evolution, environmental perturbations, or additional interconnection losses that may arise in extended hollow-core links. Second, the finite-size analysis is performed against collective attacks with a projected block size of $10^{10}$ symbols, without dedicated optimizations for extended transmission distances, representing a critical area for future algorithmic refinement. A fully composable security implementation would require explicit allocation of correctness, secrecy, parameter-estimation, error-verification, and privacy-amplification failure probabilities, alongside a rigorous extension to coherent attacks. Real-time deployment will require integrated coherent receivers characterized by high bandwidths and elevated detection efficiencies, dynamic shot-noise monitoring modules, and hardware-efficient reconciliation and privacy amplification architectures capable of operating sustainably at multi-gigabaud streaming rates.[33]. In addition, the adoption of higher-order discrete constellations or Gaussian modulation offers a viable pathway to enhance the attainable SKR[9,27]. However, such implementations will impose stricter requirements on transmitter resolution, reconciliation efficiency, and security analysis. Furthermore, the overall SKR is profoundly influenced by interconnection and coupling insertion losses. Given the ultra-low intrinsic propagation loss of advanced HCFs, the system performance becomes significantly more sensitive to these discrete insertion losses, necessitating minimized connection impairments in future network integration. Finally, the present experiment is restricted to the C-band. The measured attenuation minimum near 1565 nm suggests that extending the classical payload into the L-band, and potentially exploiting other low-loss HCF transmission windows, could further increase classical capacity while maintaining spectral separation from the quantum channel.

In summary, this work shows that hollow-core transmission can reshape the operating landscape of shared-fiber quantum-secured communication by acting directly on the physical origin of coexistence noise. By combining an ultralow-loss AR-HCF link with residual-carrier-assisted phase recovery, we demonstrate simultaneous high-rate DM-CV-QKD and multi-terabit coherent classical transmission in a single fiber. The measured tolerance to classical launch power, together with positive finite-size secret-key projections at a 100-km-equivalent condition, indicates that AR-HCF can move quantum–classical coexistence beyond the noise–power trade-off imposed by conventional solid-core links. These results point to a physical-layer route for integrating high-rate CV-QKD with high-capacity coherent optical networks in future metro and inter-data-centre communication systems.

**Methods**

**Digital signal processing**

In this experiment, the digital signal processing (DSP) workflow initiated at the transmitter infrastructure prepares the high-rate discrete-modulation quantum states before electro-optic conversion. Independent dual-polarization 15-Gbaud symbol sequences are drawn from a probabilistically shaped 16-QAM alphabet. For m = 4 bits per symbol, the symbol probabilities approximate a two-dimensional Gaussian distribution truncated at a normalized radius κ = 3 in the in-phase/quadrature plane. The sequences are pulse-shaped with a root-raised-cosine (RRC) filter spanning 200 symbols and having a roll-off factor of β=0.1. The complete DSP routine of the transmitter is described through the following chronological operations. (i) To achieve frame synchronization, a joint dual-polarization cross-correlation is performed between the received complex fields and a local pulse-shaped template of the transmitted sequence, utilizing the unique randomness of the quantum payload to yield a single, unambiguous correlation peak for precise boundary detection. (ii) Subsequently, a dedicated training sequence (TS) characterized by a strict frame-structure ratio of 1/5 is periodically interleaved into the symbol stream to provide auxiliary channel equalization and polarization demultiplexing at the receiver side, a specialized configuration incorporated specifically to eliminate the necessity to consume the digital resolution or dynamic range of the arbitrary waveform generator (AWG). (iii) To spectrally separate the quantum signal from the residual optical carrier, the resampled X- and Y-polarization waveforms, $X_{out}$ [n] and $Y_{out}$ [n], are digitally shifted by $F_{shift}$ = 11.25 GHz before driving the dual-polarization IQ modulator. The modulation is expressed as:[26]

$$X_{shifted}[n] = X_{out}[n] \cdot \exp(-j2\pi \frac{F_{shift}}{F_s} n) \tag{5}$$

$$Y_{shifted}[n] = Y_{out}[n] \cdot \exp(j2\pi \frac{F_{shift}}{F_s} n) \tag{6}$$

This arrangement places the shaped quantum sideband away from the residual carrier at the optical-carrier frequency, reducing contamination by low-frequency electrical noise and enabling independent digital extraction of the carrier and signal at the receiver.

Upon transmission through the AR-HCF channel, the mixed optical fields undergo coherent heterodyne detection and advanced digital demodulation. To effectively evade the intensive low-frequency and 1/f electrical noise impairments inherent in baseband electronics, the local oscillator utilized at the receiver side is intentionally offset in the optical domain relative to the transmitter master laser, forcing the unshifted residual carrier to beat with the offset local oscillator upon mixing within the integrated coherent receiver, which directly generates an optical pilot tone mapped at a 1-GHz IF in the electrical domain. The four resulting electrical quadrature outputs are digitized and sampled by a high-speed real-time oscilloscope at an upgraded sampling rate of 50 GSa/s, running under a shared reference clock synchronized to the transmitter infrastructure. The complete receiver DSP routine is executed through the following sequential stages. (i) DC offsets are first removed from the sampled waveforms to eliminate baseline drift. (ii) To isolate and track the fast-drift phase noise and residual frequency offset simultaneously, a digital matched filter matched to the pilot wave shape is applied to precisely extract the complex time-domain envelope of the residual carrier component. The baseline frequency offset and the continuous phase fluctuations across both polarization states are then systematically eliminated by directly multiplying the digitized quantum signal by the complex conjugate of the extracted residual carrier envelope, effectively achieving self-heterodyne-equivalent phase and frequency compensation. (iii) Following this carrier recovery, the quantum sidebands are shifted back to the baseband by -11.25 GHz through digital exponential frequency conversion and resampled to 2. (iv) A Gram–Schmidt orthogonalization procedure is performed next to compensate for any residual quadrature IQ imbalance introduced by the coherent receiver optics. (v) The digitized baseband waveforms then pass through a dedicated frame synchronization block, where a joint dual-polarization cross-correlation algorithm is performed against the local pulse-shaped template of the transmitted sequence to lock the precise frame boundaries under extreme low-SNR conditions. (vi) Polarization demultiplexing and residual dispersion compensation are initially assisted by a 71-tap, 2 × 2 multiple-input multiple-output (MIMO) equalizer executing the radius-directed algorithm (RDE). (vii) To track residual phase noise and mitigate inter-symbol interference under extreme low-SNR conditions without sacrificing the digital resolution or dynamic range of the digital-to-analog converters DACs, a superimposed TS-aided LMS equalization framework is integrated into the receiver DSP pipeline. At the transmitter, known pseudo-random training sequences are directly superimposed onto the quantum payload symbols in the amplitude domain as $x(n) = s(n) + \alpha TS(n)$, where the power scaling factor $\alpha$ is configured at a suppressed ratio of 1/5. At the receiver, a sliding-window averaging mechanism is applied across n = 16 continuous overlapping blocks. Through this long-term time-domain averaging, the statistically independent additive white Gaussian noise (AWGN) and payload data are effectively canceled out, isolating a high-fidelity, noise-suppressed training reference. This refined reference is subsequently utilized to drive a 31-tap linear butterfly-structured equalizer, where the tap coefficient matrix W is dynamically updated via the data-aided LMS stochastic gradient descent algorithm: $W(n+1) = W(n) + \mu \cdot e(n) \cdot U^{*}(n)$. Finally, the phase-compensated signals are downsampled to one sample per symbol for subsequent constellation restoration and secure key distillation.[34]

**Noise calibration**

Accurate SKR estimation requires all receiver variances to be calibrated in shot-noise units (SNU). Because the quantum symbols pass through adaptive DSP before the modulation variance $V_B$ is evaluated, the shot-noise and electronic-noise records are processed by the same DSP chain with all adaptive parameters frozen. With the signal from the transmitter side blocked and the local oscillator active, the total post-DSP variance is measured at the receiver side, which contains shot noise and electronic noise.[35]

With Alice's signal blocked and the local oscillator active, Bob measures the total post-DSP variance $\sigma^2_{tot}$, which contains shot noise and electronic noise. With both signal and local-oscillator paths blocked, the electronic-noise variance $\sigma^2_{el}$ is measured. The digital shot-noise reference for each electrical quadrature channel is then $\sigma^2_{shot} = \sigma^2_{tot} - \sigma^2_{el}$.

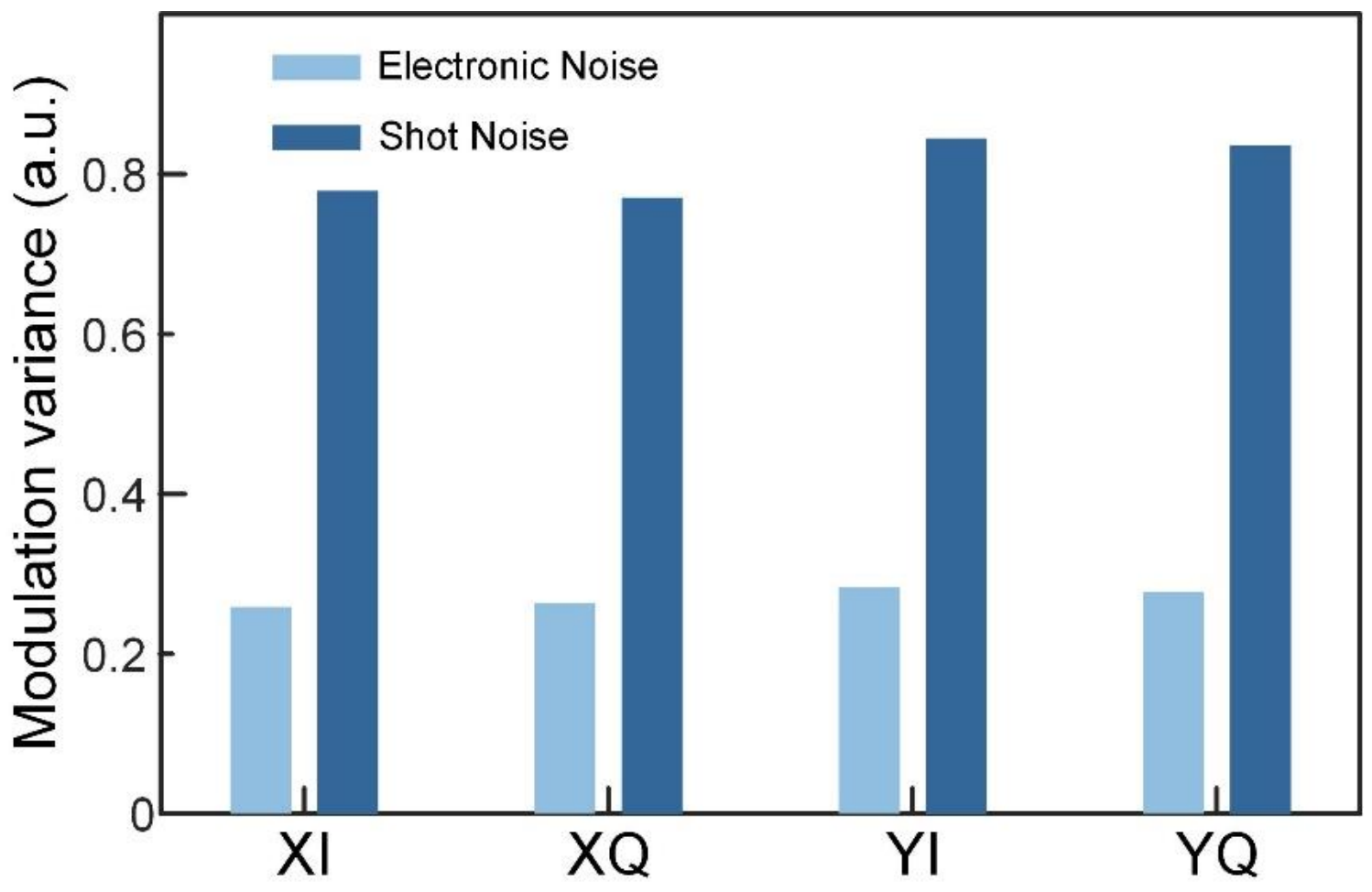

**Fig. 8 | Noise calibration for the integrated coherent receiver.** Signal-derived DSP parameters are frozen and applied identically to the signal, shot-noise, and electronic-noise records for each electrical quadrature output.

The Gram–Schmidt matrices, power-normalization factors, polarization-equalizer taps and residual-carrier phase estimates obtained from the active signal record are applied unchanged to the corresponding vacuum and electronic-noise records. Fig. 8 illustrates this calibration result. The post-DSP noise variance for each receiver output is expressed as:

$$v_{el} = \frac{\sum_i \sigma_{el(i)}^2}{\sum_i \sigma_{tol(i)}^2 - \sum_i \sigma_{el(i)}^2} \tag{7}$$

where i indexes the in-phase or quadrature output of the coherent receiver. This procedure places the recovered quantum symbols and all noise terms on a common SNU scale.

**Post-processing for CV-QKD**

Reverse reconciliation is implemented with multidimensional mapping and MET-LDPC decoding. The continuous variables originate from the probabilistically shaped 16-QAM DM CV-QKD sequence. Eight-dimensional reconciliation based on Householder transformations maps correlated Gaussian variables to binary-input channels, after which log-likelihood-ratio belief propagation is applied to the MET-LDPC matrices. The processing block size used for the reconciliation implementation is $8^{36}$.

Four seed matrices are used for the investigated loss conditions, with code rates of 0.2, 0.1, 0.05, and 0.02 for 24.3, 50, 75, and 100 km, respectively (Table 3).[17,18] Rate adaptation is achieved by puncturing or shortening around each seed rate. Because extensive puncturing and shortening degrade decoding thresholds, separate seed matrices are used for different SNR ranges. The resulting codes retain reconciliation efficiencies above 90% over their target operating regions and reach a maximum efficiency above 98%.

| Distance (km) | Code rate | Degree distribution | σ* |
|---|---|---|---|
| 24.3 | 0.2 | $\begin{cases} v(r,x) = 0.1292 r_1 x_1^3 x_2^{10} + 0.1315 r_1 x_1^2 x_2^7 + 0.7393 r_1 x_3 \\ u(x) = 0.0109 x_1^{14} + 0.0498 x_1^{10} + 0.0054 x_2^2 x_3 + 0.7339 x_2^3 x_3 \end{cases}$ | 1.712 |

| 50 | 0.1 | $\begin{cases} v(r,x) = 0.1 r_1 x_1^3 x_2^{20} + 0.025 r_1 x_1^3 x_2^{25} + 0.875 r_1 x_3 \\ u(x) = 0.025 x_1^{15} + 0.875 x_2^3 x_3 \end{cases}$ | 2.535 |
|---|---|---|---|
| 75 | 0.05 | $\begin{cases} v(r,x) = 0.04 r_1 x_1^2 x_2^{34} + 0.03 r_1 x_1^3 x_2^{34} + 0.93 r_1 x_3 \\ u(x) = 0.01 x_1^8 + 0.01 x_1^9 + 0.41 x_2^2 x_3 + 0.52 x_2^3 x_3 \end{cases}$ | 3.674 |
| 100 | 0.02 | $\begin{cases} v(r,x) = 0.0225 r_1 x_1^2 x_2^{57} + 0.0175 r_1 x_1^3 x_2^{57} + 0.96 r_1 x_3 \\ u(x) = 0.01065 x_1^3 + 0.00935 x_1^7 + 0.6 x_2^2 x_3 + 0.36 x_2^3 x_3 \end{cases}$ | 5.91 |

**Table 3 | MET-LDPC degree distributions and decoding thresholds.**

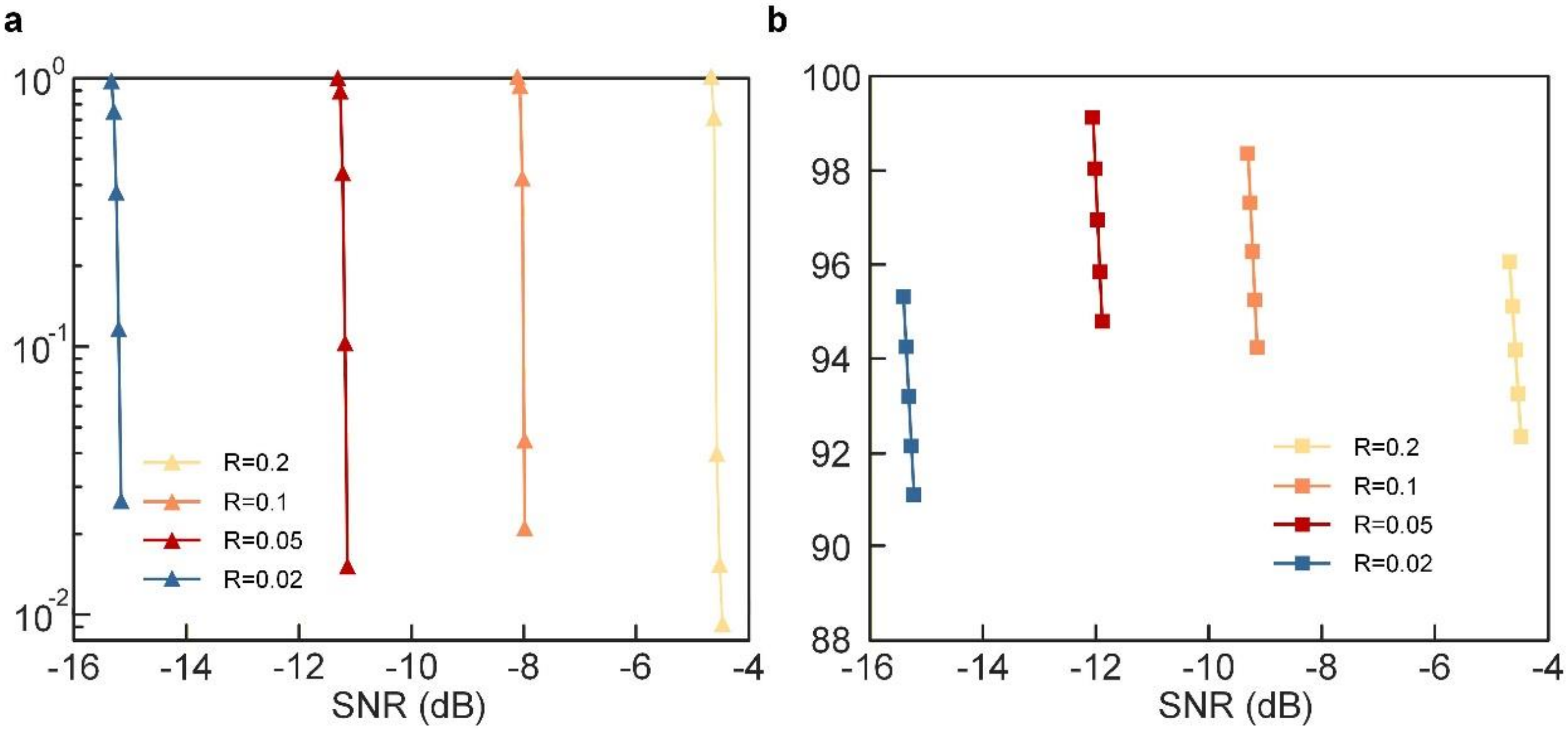

**Fig. 9 | Performance of the MET-LDPC reconciliation codes. (a)** FER versus SNR. **(b)** Reconciliation efficiency versus SNR. The lifting sizes are 50 for code rates R = 0.2, 0.1, 0.05, and 0.02, respectively.

Fig. 9 shows the FER and reconciliation efficiency of the four matrices as functions of SNR. The measured operating SNRs fall within the waterfall regions of the corresponding codes, enabling the distance-dependent rate adaptation used in the SKR calculation.

## Data availability

The data supporting the findings of this study are available within the Article and its Supplementary Information and from the corresponding authors upon reasonable request.

## Code availability

The code supporting the data processing and analysis will be made available in a public repository upon publication.

### Acknowledgements

This work was supported by the National Key Research and Development Program of China (No. 2025YFA1212900), the National Natural Science Foundation of China (No. 62225110, 62175079 and 62205119), and Hubei Province Technological Innovation Program Project (2025BAA003) and Hubei Optical Fundamental Research Centre. The authors thank Eleni Diamanti and Baptiste Gouraud for their valuable discussions and constructive comments on the manuscript.

### Author contributions

The concept of this work was conceived by S.Y. and M.T. X.J. and S.C. developed the experimental systems, performed the measurements, and analysed the data with the assistance of Z.M. P.L. contributed the hollow-core fiber and its characterization. Y.C. and J.G. contributed to system implementation, measurements and data analysis. R.L. and D.B. contributed to the security analysis and interpretation of the results. S.Y. and M.T. supervised the project. X.J., S.C., and S.Y. wrote the manuscript with input from all authors.

### Competing interests

The authors declare no competing interests.

### Additional information

**Supplementary information** is available for this paper.

### References

1. Paraïso, T. K. *et al.* A photonic integrated quantum secure communication system. *Nat. Photonics.* **15**, 850–856 (2021).
2. Dou, T. *et al.* 40 Tbps classical communication coexistence with quantum key distribution over hundred-kilometer hollow-core fiber. *Photon. Res.* **14**, 1658 (2026).
3. Wu, Q. *et al.* Integration of quantum key distribution and high-throughput classical communications in field-deployed multi-core fibers. *Light Sci. Appl.* **14**, 274 (2025).
4. Ji, X. *et al.* Quantum-secured DSP-lite data transmission architecture for AI-driven data centers. *Adv. Photon.* **7**, (2025).
5. Zhong, H.-S. *et al.* Quantum computational advantage using photons. *Science* **370**, 1460–1463 (2020).

6. Fang, L. Optical logic array: a photonic solution towards universal computing. *Front. Optoelectron.* **17**, 40 (2024).
7. Lin, X. *et al.* Polarization and wavelength routers based on diffractive neural network. *Front. Optoelectron.* **17**, 22 (2024).
8. Jouguet, P., Kunz-Jacques, S., Leverrier, A., Grangier, P. & Diamanti, E. Experimental demonstration of long-distance continuous-variable quantum key distribution. *Nat. Photonics* **7**, 378–381 (2013).
9. Hajomer, A. A. E. *et al.* Continuous-variable quantum key distribution at 10 GBaud using an integrated photonic-electronic receiver. *Optica,* **11**, 1197–1204 (2024).
10. Wang, H. *et al.* High-rate continuous-variable quantum key distribution over 100 km fiber with composable security. *Optica,* **12**, 1657–1667 (2025).
11. Hajomer, A. A. E., Derkach, I., Usenko, V. C., Andersen, U. L. & Gehring, T. Coexistence of Continuous-Variable Quantum Key Distribution and Classical Data over 120 km Fiber. *Phys. Rev. Lett.* **135**, 170804 (2025).
12. Iqbal, M. *et al.* SDN-enabled CV-QKD and classical channels coexistence: key insights for dense wavelength division multiplexing. *J. Opt. Commun. Netw.,* **17**, B28–B37 (2025).
13. Petrovich, M. *et al.* Broadband optical fibre with an attenuation lower than 0.1 decibel per kilometre. *Nat. Photonics.* **19**, 1203–1208 (2025).
14. Poletti, F. Nested antiresonant nodeless hollow core fiber. *Opt. Express,* **22**, 23807–23828 (2014).
15. Chen, S. *et al.* Coherent optical interconnects using Fermat number transform and hollow core fibre. *Commun. Eng.* **4**, 169 (2025).
16. Wang, X., Zhang, Y., Yu, S. & Guo, H. High efficiency postprocessing for continuous-variable quantum key distribution: using all raw keys for parameter estimation and key extraction. *Quantum Inf Process* **18**, 264 (2019).
17. Wang, X., Zhang, Y.-C., Li, Z., Xu, B., Yu, S. & Guo, H. Efficient rate-adaptive reconciliation for continuous-variable quantum key distribution. *Quantum Inf. Comput.* **17**, 1123–1134 (2017).
18. Zhang, K. *et al.* Effective rate-adaptive reconciliation for CV-QKD using QC-MET-LDPC codes. *Sci. China Inf. Sci.* **68**, 180510 (2025).
19. Habib, M. S., Antonio-Lopez, J. E., Markos, C., Schülzgen, A. & Amezcua-Correa, R. Single-mode, low loss hollow-core anti-resonant fiber designs. *Opt. Express,* **27**, 3824–3836 (2019).
20. Peters, N. A. *et al.* Dense wavelength multiplexing of 1550 nm QKD with strong classical channels in reconfigurable networking environments. *New J. Phys.* **11**, 045012 (2009).
21. Hua, X. *et al.* Coexistence of silicon-photonics-based QKD with classical communication over a hollow-core fiber. *Photon. Res.* **14**, 2469 (2026).
22. Fang, X. *et al.* Overcoming laser phase noise for low-cost coherent optical communication. *Nat Commun* **15**, 6339 (2024).
23. Guo, Y. *et al.* Minimalist terahertz wireless transceiver in integrated photonics. *Nat Commun.* **17**, 4429 (2026).
24. Zhu, G. *et al.* Integrated sensing and communication system exceeding a 200 km repeaterless fiber link. *Photon. Res.* **14**, 1984 (2026).
25. Marie, A. & Alléaume, R. Self-coherent phase reference sharing for continuous-variable quantum key distribution. *Phys. Rev. A* **95**, 012316 (2017).
26. Wen, H. *et al.* 16.85 Tb/s Coherent Communications Enabled by On-Chip Broadband Electro-Optic Frequency Combs. *ACS Photonics* **13**, 2134–2140 (2026).
27. Hajomer, A. A. E. *et al.* Chip-based 16 GBaud continuous-variable quantum key distribution. *Optica,* **13**, 1035–1042 (2026).
28. Tian, Y. *et al.* High-performance long-distance discrete-modulation continuous-variable quantum key distribution. *Opt. Lett.,* **48**, 2953–2956 (2023).

29. Laudenbach, F. *et al.* Continuous-Variable Quantum Key Distribution with Gaussian Modulation—The Theory of Practical Implementations. *Advanced Quantum Technologies* **1**, 1800011 (2018).
30. Ruppert, L., Usenko, V. C. & Filip, R. Long-distance continuous-variable quantum key distribution with efficient channel estimation. *Phys. Rev. A* **90**, 062310 (2014).
31. Leverrier, A., Grosshans, F. & Grangier, P. Finite-size analysis of a continuous-variable quantum key distribution. *Phys. Rev. A* **81**, 062343 (2010).
32. Kong, W. *et al.* Experimental coexistence of quantum key distribution and high-power classical communication over 101.6 km hollow-core fiber. *J. Opt. Commun. Netw.,* **17**, 914–924 (2025).
33. Hajomer, A. A. E. *et al.* Experimental composable key distribution using discrete-modulated continuous variable quantum cryptography. *Light Sci. Appl.* **14**, 255 (2025).
34. Chen, Y. *et al.* All-optical polarization split of the signal and LO for a bi-directional self-homodyne coherent system. *Opt. Lett.,* **46**, 2819–2822 (2021).
35. Roumestan, F. *et al.* Shaped Constellation Continuous Variable Quantum Key Distribution: Concepts, Methods and Experimental Validation. *Journal of Lightwave Technology* **42**, 5182–5189 (2024).
36. Zhang C M, Li M, Huang J Z, et al. Fast implementation of length-adaptive privacy amplification in quantum key distribution. *Chinese Phys. B* **23**, 090310 (2014).